\documentclass{moriond}

\newcommand{\invisible}[1]{}

\newcommand{\rien}{}

\newcommand{\INVISIBLE}[1]{}

\newcommand{\formerlongtext}[1]{
}

\newcommand{\dzero}  {D0}

\newcommand{\sqrts}{\ensuremath{\rien{\sqrt{s}}}}

\newcommand{\gev}  {\ensuremath{\mathrm{ GeV}}}
\newcommand{\tev}  {\ensuremath{\mathrm{ TeV}}}

\newcommand{\fbinv}{\ensuremath{\mathrm{ fb^{-1}}}}

\newcommand{\pt}{{\ensuremath{ p_T}}}

\newcommand{\ppbar}  {\ensuremath{p\bar p}}

\newcommand{\ttbar}  {\ensuremath{t\bar t}}

\newcommand{\BEA}{\begin{eqnarray}}
\newcommand{\EEA}{\end{eqnarray}}

\def\beq  {\begin{equation}}
\def\eeq  {\end{equation}}

\renewcommand{\dzero}{D0}
\newcommand{\invfb}{\fbinv}

\newcommand{\GeV}{\ensuremath{\mathrm{GeV}}}

\def\be{\begin{equation}}
\def\ee{\end{equation}}
\def\bea{\begin{eqnarray}}
\def\eea{\end{eqnarray}}

\newcommand{\Photo}{\includegraphics[height=35mm]{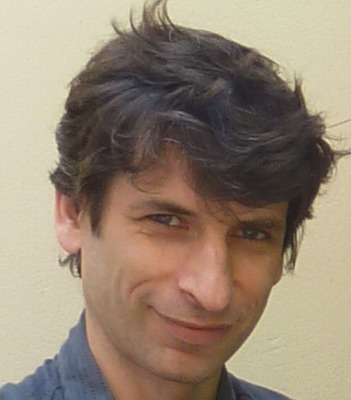}}

\begin{document}
\vspace*{4cm}
\title{Recent Top quark results from the Tevatron}

\author{Boris Tuchming  \\(for the CDF and \dzero\ Collaborations)}

\address{ Irfu/DPhp, CEA Saclay,\\ 91191 Gif-Sur-Yvette,  France}

\maketitle
\abstracts{
  We present recent measurements on top quark physics obtained at the Fermilab Tevatron $p\bar p$ collider by the CDF and D0 collaborations.
  We discuss the measurements of the top quark mass using direct and indirect methods of extraction, the forward-backward production asymmetries, and the top quark polarization.
}

\section{Introduction}
During the Run II~(2001-2011) of the Tevatron $\ppbar$ collider at $\sqrts=1.96~\tev$,  approximately ~10~\fbinv\ of data has been recorded in the \dzero\ and CDF experiments, allowing  to precisely study the top quark properties from  $\ppbar\to\ttbar$ processes.
In the following, we summarize  latest legacy results obtained either by D0 or CDF, or by combining experimental data from both experiments. In a first part we present measurements of the top quark mass, obtained either directly, or indirectly from the cross-section measurements. We then present the combination of forward-backward production asymmetries, and we conclude with the latest measurement of the top quark polarization obtained at CDF using dilepton events.

\section{Top quark mass}
The top quark  with a mass of ~$\simeq 175$~GeV  is the heaviest known fundamental particle of the Standard Model (SM). 
 Precise measurements of its mass are inputs for consistency checks of the SM and could also determine whether the electroweak vacuum is unstable,  metastable, or stable~\cite{Degrassi:2012ry}.

\subsection{D0 combination}

D0 has recently produced its legacy combined result regarding direct measurements of the top quark mass~\cite{Abazov:2017ktz}.
 The combination  is based on measurements in the lepton+jets and dilepton channels,
 using all the data collected in Run~I (1992--1996) and  Run~II (2001--2011),
corresponding to integrated luminosities of 0.1~\invfb\ and 9.7~\invfb, respectively. 
The combined mass, calculated with the ``best linear unbiased estimate'' (BLUE) method, is 
$m_t=174.95\pm 0.40 \mbox{ (stat)} \pm 0.64 \mbox{ (syst)}=174.95\pm0.75~\GeV$, which has a 0.43\% relative uncertainty.
This accuracy is much better than what had been anticipated at the beginning of Run~II.
The dominant uncertainty sources  are the statistical uncertainty, the jet energy scale (JES) uncertainty arising from the lepton+jets calibration (0.41~\gev, also of statistical origin), and the signal modeling (0.35~\gev).


\subsection{Tevatron combination}
\begin{figure}[hbt]
 
\begin{minipage}{0.5\linewidth}
\centerline{
\includegraphics[height=0.45\textheight]{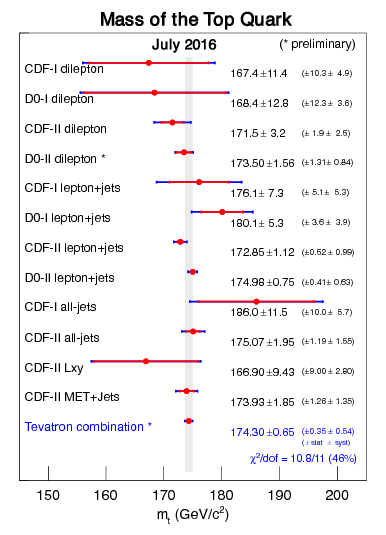}  }
\end{minipage}
\hfill
\begin{minipage}{0.5\linewidth}
    \centerline{\
\includegraphics[height=0.45\textheight]{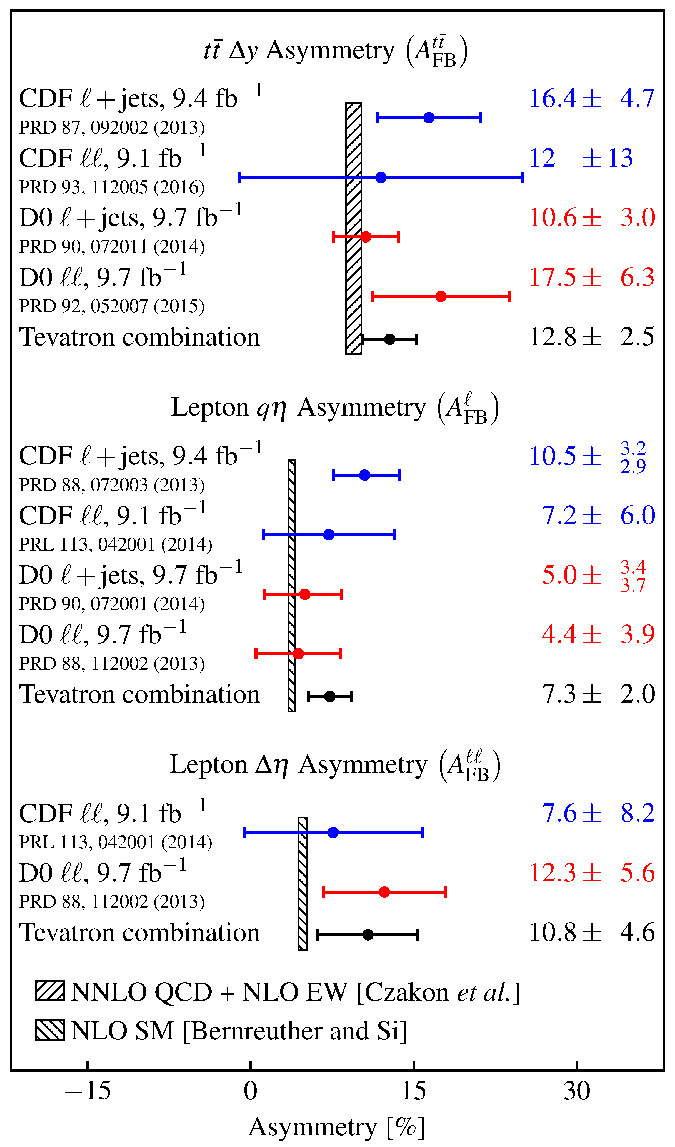}   }
\end{minipage}
\caption{Summary of Tevatron top quark mass (left) and $\ppbar\to\ttbar$ forward-backward asymmetry (right) measurements and their combinations. \label{fig:tevmass}\label{fig:tevafb}}
\end{figure}

The combined Tevatron top quark mass~\cite{Tevatron:2016combo} is based on the most recent published
measurements from both D0 and CDF. All uncertainties and their correlations are accounted for using the BLUE method.
As for the D0 measurement, the dominant sources of uncertainty are the statistical uncertainty, the JES uncertainty arising from the lepton+jets calibration (0.31~\gev), and the signal modeling (0.36~\gev).
The combined mass is
$m_t=174.30\pm 0.35\mbox{ (stat)} \pm 0.54 \mbox{ (syst)}~\GeV$, which has a 0.37\% relative uncertainty.
The summary of Tevatron measurements is shown in Fig.~\ref{fig:tevmass}.


\subsection{Indirect extraction of the pole mass from cross section}
A first measurement by D0 exploits the dependence of the theoretical prediction of the inclusive
$\ttbar$ cross section as a function of the top quark pole mass~\cite{massxs}. The inclusive cross section is measured using the full Run~II dataset. The top quark mass is extracted  accounting
for the dependence of the measurement as a function of the mass due to acceptance effects and using the  next-to-next-to-leading-order 
(NNLO) cross-section calculation of  top++~\cite{top++}. D0 achieves a 1.9\% precision:
$m_t=172.8\pm 1.1 \mbox{ (theory)} ^{+3.3}_{-3.1} \mbox{ (experimental)}~\gev$.

More recently, D0 performs another extraction using  its measured  differential $\ttbar$ cross section~\cite{Abazov:2014vga} as a function of sensitive kinematic variables: the mass of the $t\bar t$ system ($m(\ttbar)$) and the transverse momentum of the top quarks ($\pt(t/\bar t)$).
The mass is extracted from a fit of either the NLO or the NNLO cross section~\cite{Czakon:2016ckf} to the unfolded D0 data. The mass is extracted with different factorization and renormalization scales and different PDF to obtain the theory uncertainty.
The measurement using NNLO calculations is
$m_t=169.1 \pm 1.1 \mbox{ (theory)} \pm 2.2 \mbox{ (experimental)}~\gev$ which achieves
a 1.5\% accuracy~\cite{D0conf6473} and  represents a 25\% improvement over the measurement using the inclusive cross-section, thus demonstrating the power of the method.

\section{Forward-backward production asymmetries}

\def\yt {\ensuremath{ y_t}}
\def\ytbar {\ensuremath{ y_{\bar t}}}
\def\dyttbar{\ensuremath{\Delta y_{\ttbar}}}
\newcommand{\defAll}    {\ensuremath{\All = \frac{N(\Delta\eta > 0) - N(\Delta\eta < 0)}{N(\Delta\eta > 0) + N(\Delta\eta < 0)}}}
\newcommand{\Alfb}      {\ensuremath{A^{\ell}_{\rm FB}}}
\newcommand{\defAlfb}   {\ensuremath{\Alfb = \frac{N_\ell(q\times \eta>0) - N_\ell(q\times \eta<0)}{N_\ell(q\times \eta>0) + N_\ell(q\times \eta<0)}}}
\newcommand{\Attfb}       {\ensuremath{A^{t\bar t}_{\rm FB}}}
\newcommand{\Att}       {\Attfb}
\newcommand{\defAtt}{\ensuremath{\Att=   \frac {  N ( \dyttbar>0) -  N ( \dyttbar <0) }{  N ( \dyttbar>0) + N ( \dyttbar <0) }}}
\newcommand{\All}       {\ensuremath{A^{\ell\ell}_{\rm FB}}}

The SM predicts a small positive forward-backward asymmetry in the $\ppbar\to\ttbar$ production due to both
strong (QCD) and electroweak (EW) quantum effects: the (anti)top quarks tend to go in the same direction as the  incoming (anti)protons.
When the first measurements were performed at Tevatron Run~II,  large departures from  SM expectations were observed which brought a
lot of excitement~\cite{Abazov:2007ab,Aaltonen:2008hc,Aaltonen:2011kc,Abazov:2011rq} as potential indication of new physics beyond the SM.
Since then,  the theory prediction has been improved by including higher order QCD and EW corrections~\cite{Czakon:2014xsa},
and both CDF and D0 have completed their asymmetry measurement program using dilepton and lepton+jets channels in the full Run~II dataset.
They now provide the combination of their measurements obtained with the BLUE method, accounting for all uncertainties and their correlations~\cite{TevconfAFB}.

The combinations are performed for the three inclusive asymmetries,  \defAtt, \defAlfb, and \defAll, based either on the reconstructed rapidities of the top and antitop ($\Delta y=y_t-y_{\bar t}$)  or on the lepton pseudorapidities ($\eta$), or on the difference between lepton pseudorapidities for dilepton channels ($\Delta\eta=\eta_{\ell^+}-\eta_{\ell^-}$). Here, each $N$ represents  a number of events corrected for acceptance and efficiency effects.
The combination of differential measurements of \Attfb\ is also performed as a function of $\Delta y$ and of the $\ttbar$ mass $(m(\ttbar))$.
As the different measurements are dominated by the uncorrelated statistical uncertainties,
the overall correlations are small, at the level of 10\%, mainly due to signal modeling uncertainty. However, large correlation effects due to data unfolding have to be properly taken into account for the differential combinations.

\begin{figure}[htb]\center
\includegraphics[width=0.45\textwidth]{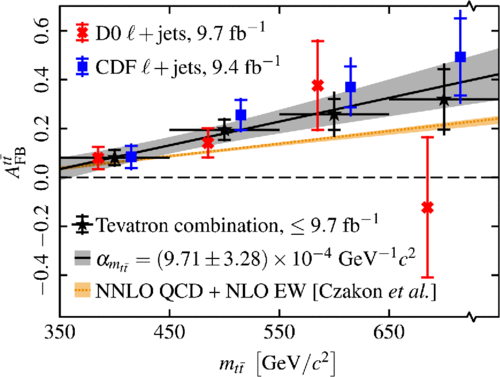} \hfill
\includegraphics[width=0.45\textwidth]{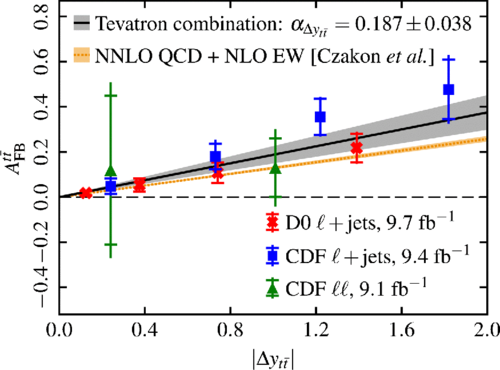}   
\caption{Differential forward-backward asymmetry as a function of $m(\ttbar)$ and $\Delta y$.\label{fig:TevDiffAFB}}
\end{figure}

In Fig.~\ref{fig:tevafb} we summarize the combined inclusive measurements for the three asymmetries.
The differential asymmetries are shown in Fig.~\ref{fig:TevDiffAFB}, where linear fits are used to model the combined asymmetries.
All combined differential and inclusive asymmetries are compatible within 1.3--1.6 standard deviation of the NNLO QCD + NLO EW  predictions of the SM.

\section{Top quark polarization}
CDF has recently released a new measurement of top quark polarization using dilepton events~\cite{Cdfpolar}.
This measurement is performed in the ``transverse'' and ``helicity'' basis, where the main axes are, in the $\ttbar$ zero momentum frame, the axis perpendicular
to the production (top,beam) plane, and the $\ttbar$ axis, respectively.
This new measurement completes previous measurements at D0 in the dilepton channel
(beam basis)~\cite{Abazov:2015fna} and lepton+jets channel (beam, transverse, and helicity bases)~\cite{Abazov:2016tba}.
The observables are given by the lepton angles  ($\theta^{\pm}$) relative to the reference axis, where the reconstructed lepton directions are obtained after successively boosting the leptons in the \ttbar\ rest frame and in the parent top quark rest frame. The differential angular distribution for $\ell^+$ and $\ell^-$ follows the relation
$\frac 1 {\sigma} \frac {d^2\sigma}{ d\cos\theta^+d\cos\theta^-}=
\frac 1 4
( 1+  \alpha^+P^+\cos\theta^+ +  \alpha^-P^-\cos\theta^- - C \cos\theta^+\cos\theta^-)
$,
where $ \alpha$ is the spin analyzing power of the particle ($\simeq 0.99$ within the SM),  $ P^{\pm}$ is the polarization, and $C$ is the spin correlation.
CDF measures the polarization using two-dimensional templates in $(\cos\theta^+,\cos\theta^-)$,
considering two scenarios,
one with CP conservation ($\alpha^+ P^+=\alpha^-P^-$),
and the other one with maximal CP violation ($\alpha^+ P^+=-\alpha^-P^-$). The results are 
$ \alpha P^{\mathrm {CPC}}_\mathrm{helicity} = -0.130 \, ^{+0.114}_{-0.109}\,  (\mathrm{stat})\, \pm 0.111 \, (\mathrm{syst}),$
$ \alpha P^{\mathrm {CPV}}_\mathrm{helicity} = -0.046\,  \pm 0.123\,  (\mathrm{stat})\, \pm 0.040 \, (\mathrm{syst}),$
$ \alpha P^{\mathrm {CPC}}_\mathrm{transverse} = -0.077\,  \pm 0.177\,  (\mathrm{stat})\, \pm 0.098 \, (\mathrm{syst}),$ and
$\alpha P^{\mathrm {CPV}}_\mathrm{transverse} = -0.111\, ^{+0.114}_{-0.109}\,  (\mathrm{stat})\, ^{+0.055}_{-0.056} \, (\mathrm{syst}).
$
The results  as well as the previous D0 results are compatible with the SM which predicts polarization below the percent level in all bases.


\section{Conclusion}

The realm of top quark physics has been pioneered at Tevatron since the top quark discovery in 1995.
The Tevatron is still producing competitive results in the finalization of its legacy measurements.
The mass of the top quark is measured with a 0.37\% accuracy by combining D0 and CDF. D0 also performs and indirect extraction of the pole mass with a 1.5\% accuracy.
The \ttbar\ differential and inclusive forward-backward asymmetries are in agreement with the Standard Model. Finally, Tevatron also measures the top quark polarization in agreement with the Standard Model.

\end{document}